\documentclass[letterpaper,twocolumn,10pt]{article}
\usepackage{usenix2019_v3}

\usepackage{hyperref}
\usepackage{tikz}
\usepackage{amsmath}

\usepackage{pifont}

\usepackage{booktabs}
\usepackage{svg}

\usepackage{dirtree,array}
\usepackage{booktabs}

\usepackage{filecontents}

\usepackage{wrapfig}
\usepackage{paralist}
\usepackage{xspace}
\usepackage{tabularx}
\usepackage{subfigure}
\usepackage{multirow}
\usepackage{enumitem}

\usepackage[utf8]{inputenc}

\usepackage[nameinlink]{cleveref}
\crefformat{section}{\S#2#1#3}
\crefformat{subsection}{\S#2#1#3}
\crefformat{subsubsection}{\S#2#1#3}

\usepackage[compact]{titlesec}

\def\notodocomments

\date{}

\newcommand{\mycaption}[2]{\caption{\textbf{#1}. {#2}} \vspace{-10pt}}
\newcommand{\sref}[1]{\S\ref{#1}}

\newcommand{\ie}{\textit{i.e.,}\xspace}

\newcommand{\myx}{$\times$\xspace}

\newcommand{\load}{\texttt{load}\xspace}
\newcommand{\store}{\texttt{store}\xspace}

\newcommand{\sysmalloc}{\texttt{malloc()}\xspace}

\newcommand{\sysfree}{\texttt{free()}\xspace}

\ifdefined\notodocomments
\newcommand{\haoran}[1]{}
\newcommand{\baris}[1]{}

\newcommand{\rohancomment}[1]{}
\newcommand{\saurabh}[1]{}
\newcommand{\mde}[1]{}
\newcommand{\gilbert}[1]{}
\newcommand{\reviewercomment}[1]{}
\providecommand{\TODO}[1]{}
\providecommand{\todo}[1]{}
\else
\newcommand{\haoran}[1]{{\textcolor{blue}{[[Haoran: #1]]}}}
\newcommand{\baris}[1]{{\textcolor{olive}{[[Baris: #1]]}}}

\newcommand{\rohancomment}[1]{{\textcolor{olive}{[[Rohan: #1]]}}}
\newcommand{\saurabh}[1]{{\textcolor{purple}{[[Saurabh: #1]]}}}
\newcommand{\mde}[1]{{\textcolor{red}{[[Mike: #1]]}}}
\newcommand{\gilbert}[1]{{\textcolor{magenta}{[[Gilbert: #1]]}}}
\newcommand{\reviewercomment}[1]{{\textcolor{purple}{[[Reviewer: #1]]}}}
\renewcommand{\reviewercomment}[1]{}
\providecommand{\TODO}[1]{{\protect\color{red}\noindent {\bf [[TODO} \emph{#1}{\bf ]]}}}
\providecommand{\todo}[1]{{\protect\color{red}\noindent {\bf [[TODO} \emph{#1}{\bf ]]}}}
\fi

\newcommand{\Jenga}{Jenga\xspace}
\newcommand{\sysname}{\Jenga\xspace}

\newcommand{\Memtis}{Memtis\xspace}
\newcommand{\memtis}{\Memtis}
\newcommand{\coolinginterval}{cooling interval\xspace}
\newcommand{\Hybridtier}{HybridTier\xspace}
\newcommand{\hybridtier}{\Hybridtier}
\newcommand{\MemtisQC}{\Memtis-QC\xspace}
\newcommand{\memtisqc}{\MemtisQC}
\newcommand{\glibc}{{glibc-malloc}\xspace}
\newcommand{\btree}{{Btree}\xspace}

\newcommand{\roms}{{654.roms\_s}\xspace}
\newcommand{\bwaves}{{603.bwaves\_s}\xspace}

\newcommand{\GAPBSBC}{GAPBS-BC\xspace}
\newcommand{\GAPBSPR}{GAPBS-PR\xspace}
\newcommand{\gapbsbc}{\GAPBSBC}
\newcommand{\gapbspr}{\GAPBSPR}

\newcommand{\numApplications}{10\xspace}
\newcommand{\numPreviousSystems}{5\xspace}

\newcommand{\jengaVsSecondBestFastTierEqualsWSS}{$\sim$28}
\newcommand{\jengaVsSecondBestFastTierLargerThanWSS}{$\sim$15}
\newcommand{\jengaVsMemtisEightGB}{50}

\def\|#1|{\mathid{#1}}
\newcommand{\mathid}[1]{\ensuremath{\mathit{#1}}}
\def\<#1>{\codeid{#1}}
\protected\def\codeid#1{\ifmmode{\mbox{\sf{#1}}}\else{\sf #1}\fi}

\renewenvironment{thebibliography}[1]{
  \begin{oldthebibliography}{#1}
    \setlength{\itemsep}{0em}
    \setlength{\parskip}{0em}
}
{
  \end{oldthebibliography}
}

\usepackage{ifthen}
\newboolean{publicversion}
\setboolean{publicversion}{false}
\ifthenelse{\boolean{publicversion}}{
  \newcommand{\grumbler}[3]{}
}{
  \newcommand{\grumbler}[3]{\textcolor{#3}{\bf #1: #2}}
}

\begin{document}

\date{}

\title{\Large \bf Jenga: Responsive Tiered Memory Management without Thrashing}

\author{
{\rm Rohan Kadekodi${^*}$}\\
University of Washington
\and
{\rm Haoran Peng${^*}$}\\
University of Washington
\and
{\rm Gilbert Bernstein}\\
University of Washington
\and
{\rm Michael D. Ernst}\\
University of Washington
\and
{\rm Baris Kasikci}\\
University of Washington
} 

\maketitle

{\renewcommand{\thefootnote}{*}
\footnotetext{Both authors contributed equally.}
}

\begin{abstract}

A heterogeneous memory has a single address space with fast access to some
addresses (a fast tier of DRAM) and slow access to other addresses (a
capacity tier of CXL-attached memory or NVM).
A tiered memory system aims to maximize the number of accesses to the
fast tier via page migrations between the fast and capacity tiers.
Unfortunately, previous tiered memory systems can perform poorly
due to (1) allocating hot and cold objects in the same page and
(2) abrupt changes in hotness measurements that lead to thrashing.

This paper presents \Jenga, a tiered memory system that addresses both
problems.
\Jenga's memory allocator uses a novel
context-based page allocation strategy.
\Jenga's accurate measurements of page hotness
enable it to react to
memory access behavior changes in a timely manner while
avoiding thrashing.
Compared to the best previous tiered
memory system,
\Jenga runs memory-intensive applications \jengaVsSecondBestFastTierEqualsWSS\% faster across 10 applications, when the fast tier capacity  matches the working set size, at a CPU overhead of <3\% of a single core and a memory overhead of <0.3\%.\looseness=-1


\label{sec:dummy-for-etags}


\end{abstract}


\section{Introduction}
\label{sec:Introduction}
Main memory cost and capacity poses a challenge in the face of data
explosion. Memory is one of the most expensive resources in datacenter
servers, amounting to 37\% of Meta's server costs~\cite{maruf2023tpp} and about 50\% of Microsoft Azure's server costs~\cite{azure50}. Furthermore, DRAM technologies
do not scale in capacity to accommodate the growing needs of modern
memory-intensive applications such as large ML models and graph processing
frameworks~\cite{epoch2023aitrends}.

In order to reduce the cost of server memory, industry is innovating on memory expanders that use cache-coherent interconnects such as
Compute eXpress Link (CXL)~\cite{cxl2022}. This enables the use of cheaper
and denser non-DRAM memory technologies such as NVM (Non-Volatile
Memory~\cite{izraelevitz2019basic}) or older generation DDR4 DRAM chips on
the PCIe5.0 bus (PCI latency is more than memory bus latency).  The memory
is directly addressable, sharing a single space with DDR5 DRAM on the
memory bus.
A heterogeneous memory contains a \emph{fast tier} of DRAM and a slower but
larger \emph{capacity tier}.  Memory tiering enables a server to access
terabytes of memory in an affordable manner \cite{tieringcost1,li2023pond,maruf2023tpp,duraisamy2023towards}.


The higher access latency of the capacity tier can significantly degrade
performance~\cite{sun2023demystifying,izraelevitz2019basic}.
In order to hide this latency,
a tiered memory system in the OS migrates
memory pages between the fast and capacity tiers.
The goal of a tiered memory system
is to maximize accesses to the fast tier (called the hit rate) while minimizing the number of page
migrations between the fast and capacity tiers.

This paper begins with a comprehensive analysis of \numApplications
widely used applications previously
  used in evaluations of tiered memory systems. We categorize these applications based on their data
allocation patterns (do they predominantly create small or large
objects?)\ and their data access behaviors (is the hot data stable or does
the hot data exhibit phase changes?). We find that previous
tiered memory
systems~\cite{lee2023memtis,maruf2023tpp,raybuck2021hemem,xiang2024nomad,ren2024mtm,duraisamy2023towards,li2023pond,yan2019nimble,yang2017autotiering,autonuma}
perform well only on applications that create large objects and where the
hot data remains stable. 
To understand this behavior, we analyzed previous tiered memory systems and
discovered two shortcomings that lead to their poor performance.

First, these systems allocate hot and cold data in the same memory page.
This can prevent all the pages with hot data from fitting in the fast tier,
resulting in a poor hit rate.

Second, current tiered memory systems are unable to accurately determine page hotness. Some systems focus solely on recent access patterns while ignoring long-term historical trends~\cite{maruf2023tpp, xiang2024nomad, autonuma,yang2017autotiering,li2023pond}, resulting in poor hit rates. Other systems attempt to handle changing workload phases by abruptly changing page hotness at fixed intervals~\cite{lee2023memtis,raybuck2021hemem,ren2024mtm,duraisamy2023towards,song2025hybridtier}, but this approach creates a performance trade-off: longer intervals cause slow adaptation to new hot data patterns, while shorter intervals trigger excessive and counterproductive data migrations as pages repeatedly move between tiers, both scenarios ultimately harming overall system performance.

This paper introduces \Jenga, which outperforms other tiered memory systems on
applications with diverse memory access behaviors. \Jenga's two main components are a \emph{context-based heap allocator} and \emph{smooth hotness tracking}.\looseness=-1

\paragraph{Context-based allocator}
\Jenga is the first tiered-memory system with a \textit{context-based
  heap allocator} for allocating an application's objects into pages in a
way that is profitable for their migration to the appropriate memory tier.


The three allocation
strategies are the traditional size-based and time-based strategies, plus a
new \textit{context-based grouping} strategy that we introduce.
Context-based allocation places objects allocated at the same stack trace and that are of the same size, 
into the same page.  Objects with the same
call stack may be semantically related and thus may be accessed together
and/or a similar number of times. For instance, a tree implementation
might have different call stacks when allocating internal nodes versus leaf
nodes. When performing tree traversals, internal nodes are visited much
more often than the leaf nodes.


\paragraph{Smooth hotness tracking}
In order to adapt to phase changes in the hot data, \Jenga introduces
\emph{smooth hotness tracking} .
Rather than periodically abruptly changing the relative hotness of pages, \Jenga applies a gradual and continuous hotness decay with every page access. Smooth hotness reduces the influence of hard-coded intervals, allowing a page's hotness to change smoothly and responsively over time. Pages from no-longer-hot regions are migrated to the capacity tier without waiting for a lengthy interval, while frequently accessed pages in the current phase retain their hotness and continue to remain in the fast tier. This smooth hotness tracking approach reduces unnecessary migrations due to thrashing, increases the hit rate by considering historical access trends and promptly responding to phase changes, and ultimately improves overall performance.
Smooth hotness tracking is inexpensive:  it requires only one additional
counter per page and little computation.

\paragraph{Our contributions}
    \textbf{Analyses:} We analyzed the behavior of previous tiered
      memory systems with \numApplications memory-intensive applications and reveal new
      findings:
      (1) allocation of objects within pages influences
      the performance of applications on tiered memory,
      and
      (2) today's tiered memory systems are not timely enough to react to some
      changes in access patterns, and they over-react to other changes.
    \textbf{\Jenga design:} Our \Jenga tiered memory system solves the
    above problems using novel context-based allocation
      and smooth hotness tracking approaches.
      \textbf{Evaluation:}
      Jenga has small CPU overhead (<3\% of a single CPU
      core) and memory overhead (<0.3\%).
      We compared \Jenga to \numPreviousSystems previous tiered-memory
      systems while varying the size of the fast tier and capacity
      tier. \Jenga runs memory-intensive applications
      \jengaVsSecondBestFastTierEqualsWSS\% faster than the best other system when the
      fast tier capacity matches the working set of the
      application. Furthermore, \Jenga is the only system that consistently achieves high performance in different applications and fast tier capacities. Among other systems, there is no single system that performs consistently well across applications and fast tier capacities. For small object applications, \Jenga
      achieves performance equal to or better than
        the best
      other system while requiring only $\frac{1}{3}$ the fast-tier
      capacity. \sysname is publicly available at \url{https://github.com/efeslab/jenga}.

\section{Background: Tiered Memory Systems}
\label{sec:Background}

A tiered memory
system~\cite{bergman2022reconsidering,dulloor2016data,gupta2015heterovisor,hildebrand2020autotm,kannan2017heteroos,kim2021exploring,li2022transparent,maruf2022multi,maruf2023tpp,raybuck2021hemem,ren2021sentinel,autonuma,tiering08,wang2019panthera,yan2019nimble,lee2023memtis,yang2017autotiering}
is responsible for managing application data such that the hot data is
stored in the fast memory tier, while the cold data is in the capacity
tier. This is performed by (1) tracking memory accesses to pages, (2)
classifying pages as \emph{hot}, \emph{warm} or \emph{cold}, and (3)
migrating pages to the appropriate memory tiers. 

\subsection{Tracking Memory Accesses}
\label{sec:background:tracking}
The memory access tracking mechanism tracks \load and \store accesses to
pages. It needs to track accesses to millions of pages with low memory and
computation overheads. Tiered memory
systems track page accesses two primary ways.

\paragraph{Counting via NUMA-hinting faults}
A kernel task periodically samples a range of a process's memory (256MB of
pages by default~\cite{autonuma}) on each NUMA node. Page access frequency
is estimated by counting the number of times the sampled NUMA page
generates a fault. Typically, this method is used for NUMA balancing
(AutoNUMA~\cite{autonuma}), but it is also used by tiered-memory systems
(e.g., AutoTiering~\cite{yang2017autotiering}, TPP~\cite{maruf2023tpp},
Tiering-0.8~\cite{tiering08}, Nomad~\cite{xiang2024nomad}). NUMA-hinting
page faults are expensive as they rely on soft page faults and TLB
invalidations to track accesses. To avoid slowing the foreground
application threads, the NUMA-hinting kernel task sacrifices accuracy
by monitoring a subset of pages during each period.


\paragraph{Counting via hardware-assisted access sampling}
Modern Intel and AMD processors include hardware support for sampling
events called Process Event-Based Sampling (PEBS) or Performance Monitoring
Unit (PMU). PEBS can notify a tiered-memory system of events such as
last-level cache misses to the fast or capacity tier. Previous tiered-memory
systems that use PEBS maintain one counter in the metadata of each page to
record the number of accesses to that page. Since PEBS does not require TLB
invalidations or page table scanning, it is scalable to larger application
working sets without incurring prohibitive memory or CPU overheads.
It has been used in recent work such as
HeMem~\cite{raybuck2021hemem}, \Memtis~\cite{lee2023memtis},
TMTS~\cite{duraisamy2023towards}, MTM~\cite{ren2024mtm},
FlexMem~\cite{xu2024flexmem} and \hybridtier~\cite{song2025hybridtier}.



\subsection{Determining Page Temperature}
\label{sec:background:hotness}
The method used for profiling memory accesses dictates how tiered-memory
systems determine page hotness.

\subsubsection{NUMA hinting uses static hot-page classification}
Due to the high cost of tracking accesses via NUMA hinting,
a page is marked hot even if it is accessed only once~\cite{autonuma,xiang2024nomad,yang2017autotiering}
or twice~\cite{maruf2023tpp,maruf2022multi}.
Thus, NUMA hinting
determines page hotness from recent accesses (not the history of
accesses), which is inaccurate~\cite{lee2023memtis,song2025hybridtier}.

\subsubsection{PEBS uses histogram-based hot-page classification}
\label{sec:background:hotness:pebs}
When using PEBS counters, the tiered memory system assigns each page to a bin
based on a weighted sum of recent and past accesses within the page.
Each bin is dynamically classified as hot, warm, or cold.

To assign pages to bins, a per-page counter tracks the number of
  accesses made to the page. A system-wide histogram tracks
  the relative hotness between pages. The histogram contains
  logarithmic-scaled bins, each bin holding a count of the number of pages
  belonging to the bin; pages in the $n$th bin have been accessed between
  $2^{n}$ and $2^{n+1}$ times (tiered memory systems typically use 16 bins).
  On every access to a page, its counter
  is incremented, and its bin index is computed.
  If its bin has changed since the previous access, the system-wide
  histogram is updated.


The tiered memory system maintains two hotness thresholds: a hot threshold
$T_{hot}$, and an optional warm threshold $T_{warm}$.\label{sec:memtis-no-warm-bin}
These thresholds are consecutive bin
indices in the system-wide histogram. 
$T_{hot}$ is set such that the total number of hot pages
$\approx$ the capacity of the fast tier, and $T_{\|warm|} = T_{\|hot|} - 1$\footnote{
When at least 75\% of the fast tier contains pages that are classified as hot (with bin index $\ge$$T_{\|hot|}$), 
\Memtis dispenses
with the warm threshold, categorizing every bin (and thus every page) as hot or cold and none as
warm.  \sref{sec:design:existing_cooling} discusses the problems that this
causes for \Memtis.}. All the bins below $T_{\|warm|}$ are categorized as cold bins.
\emph{Threshold adaptation} periodically recomputes $T_{\|hot|}$ and determines the existence of $T_{\|warm|}$ from the histogram.

If a page in the capacity tier has a bin index $\ge T_{\|hot|}$, the page
is added to the system-wide promotion queue. Similarly, if a page is the
fast tier has a bin index $\le T_{\|cold|}$, the page is added to the
system-wide demotion queue.
To reduce thrashing, warm pages are left unmigrated in the
tier they currently reside in.

PEBS-based tiered memory systems adapt to workloads that change over time
using a mechanism called \textbf{\textit{cooling}}. Cooling is a heavy-weight process that involves traversals and modifications to system-wide data structures and asynchronous modifications to the per-page access counters.


First, cooling triggers the process of decaying all access counters by a factor of $\frac{1}{2}$. Decay reduces the influence of old accesses on the hotness of a page, and therefore
increases the influence of more recent accesses. After the trigger, the first
access to a page decays its access count,
re-computes its bin index, and updates the system-wide histogram. This
\emph{asynchronous} decay of counters is done to avoid scanning the entire
memory to update page counters on each cooling
event. 

Second, cooling performs a synchronous decay of page counters for the pages in the system-wide promotion and demotion queues, to avoid unnecessary migrations.

Existing tiered-memory systems are designed for an infrequent \coolinginterval, and explicitly recommend this
  configuration to users~\cite{lee2023memtis,raybuck2021hemem}. This is intended to ensure that the system-wide histogram reflects a sufficient number of memory accesses across pages to accurately identify the hotness distribution, and also to frequently avoid costly traversals of global migration queues.
HeMem~\cite{raybuck2021hemem} triggers system-wide cooling whenever the
number of accesses for any page reaches 1024. \Memtis~\cite{lee2023memtis}
performs threshold adaptation after each 100K PEBS samples and
triggers cooling
after each 2M PEBS samples. \hybridtier~\cite{song2025hybridtier} contains two \coolinginterval{s}, called momentum interval (500K PEBS samples) and a frequency interval (80M PEBS samples), designed to capture short-term bursts as well as long-term trends. 

\subsection{Migrating Pages}
Tiered-memory systems perform migrations of the pages that are in
  the system-wide promotion and demotion queues. Background promotion and
  demotion threads periodically traverse these queues, dequeue each page,
  compare its bin index to the hotness thresholds (which may have changed
  in the meanwhile), and perform migration if necessary.
\section{Handling Memory Access Diversity}
\label{sec:analysis}

This section discusses common application memory access
patterns and how previous tiered-memory systems fare on each pattern.  

\subsection{Applications}
\label{sec:applications}
\label{sec:subject-applications}

\begin{table}[tb]
\begin{center}
\begin{tabular}{lcccl}
        & Obj. & WS & &  \\
Application & size & change & WSS & Papers \\ \hline
BTree & small & no & 8GB & \cite{lee2023memtis} \\
Silo & small & no & 8GB & \cite{raybuck2021hemem, lee2023memtis, song2025hybridtier} \\
GAPS-BC & large & yes & 4GB & \cite{lee2023memtis, raybuck2021hemem, xiang2024nomad} \\
GAPS-PR & large & yes & 4GB & \cite{lee2023memtis, xiang2024nomad} \\  
FlexKVS & large & yes & 4GB & \cite{raybuck2021hemem} \\
Liblinear & large & yes & 4GB & \cite{lee2023memtis} \\
Graph500 & large & no & 4GB & \cite{lee2023memtis} \\
XSBench & large & no & 4GB & \cite{lee2023memtis} \\
\roms & mixed & yes & 4GB & \cite{lee2023memtis, song2025hybridtier} \\
\bwaves & large & no & 4GB & \cite{lee2023memtis, song2025hybridtier} \\
\end{tabular}
\end{center}
\caption{Applications we used to evaluate memory tiering systems.
  They are the union of applications used in two previous
  evaluations~\cite{lee2023memtis, raybuck2021hemem}.
  Cherry-picking applications for evaluation would be a threat to
  external validity.}
\label{fig:subject-applications}
\end{table}

\begin{figure}
  \centering
\includegraphics[width=0.45\textwidth]{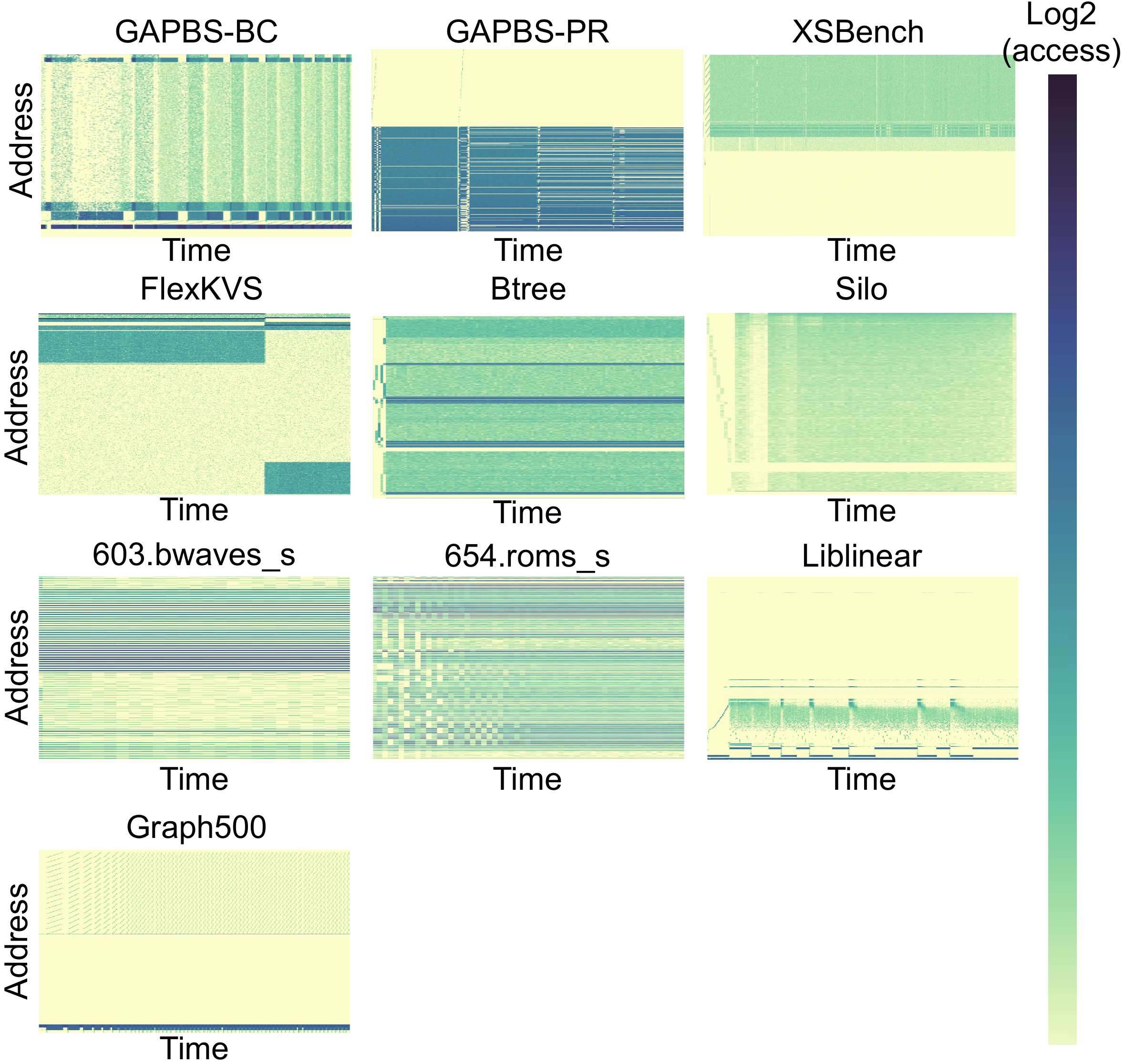}
\mycaption{Application heatmaps}{Darker colors indicate more accesses to
  a given memory range, during a given time period.
\gapbsbc, \GAPBSPR, FlexKVS, Graph500 and Liblinear create large objects and exhibit working set changes. XSBench, 603.bwaves\_s and 654.roms\_s contain a stable working set. Btree and Silo primarily allocate small objects. 
}
\label{fig:eval-heatmap-apps}

\end{figure}

\Cref{fig:subject-applications} lists the applications we used in our evaluation,
and \cref{fig:eval-heatmap-apps} shows their access heatmaps.
These applications are the union of those used in two previous evaluations
\cite{lee2023memtis, raybuck2021hemem}.
We did not cherry-pick applications for this evaluation.  Once we started to
examine a application, we retained it in our dataset regardless of whether
\Jenga worked well on it or not.
For example, \Jenga underperformed by 5--7\% on the \roms and XSBench applications.


The GAP Benchmark
Suite~\cite{beamer2015gap} is a set of graph processing applications.
Our experiments use betweenness centrality
(\GAPBSBC) and page rank (\GAPBSPR).
(The other applications in GAP have a small working set regardless of the workload.)
We evaluated \GAPBSBC on a randomly-generated undirected Kronecker Graph with $2^{29}$ vertices and an average degree of
6.
The default workload of \GAPBSPR is
highly skewed, meaning all 
hot objects fit within hundreds of MB, which is not representative
of realistic tiering scenarios.
Therefore, we run \GAPBSPR on a randomly-generated undirected
graph with $2^{30}$ vertices and average degree of 2.

XSBench~\cite{tramm2014xsbench} is the computational kernel of the Monte
Carlo neuron transport algorithm. It performs continuous macroscopic cross section lookups, and is used for performance analysis on high performance computing architectures. XSBench contains a skewed access pattern in which the working set is concentrated in the initial set of allocated
pages.

FlexKVS~\cite{kaufmann2016high} is a low-latency, scalable key--value store
used for in-memory caching compatible with Memcached.
We ran a read-only
workload on FlexKVS using its internal kvsbench
benchmark (the same benchmark used in prior work~\cite{raybuck2021hemem}), such that 90\%
of accesses occur on 20\% of elements, making up a contiguous 4GB working
set. Midway through its run, we modify the hotset (via \<kvsbench config>),
changing the hotset to a different 4GB region. This behavior is visible
in its access heatmap (\cref{fig:eval-heatmap-apps}).

The Btree key--value store~\cite{achermann2020mitosis} sets up 200M
records,
then performs 2B random reads.

Silo~\cite{tu2013speedy} is an in-memory database commonly used to evaluate
tiered memory systems.  It creates millions of objects, each tens to
hundreds of bytes in size. We ran the YCSB Run C workload (100\% reads) on
Silo with 2.5B operations on 70M keys. We report the performance of loading the database, and for servicing the reads.


Liblinear~\cite{fan2008liblinear} is an open-source library for large-scale
linear classification, and supports logistic regression and linear
support vector machines. Its workload is the KDD12 dataset.

Graph500~\cite{murphy2010introducing} is graph processing
benchmark, similar to GAP. The workload builds an
undirected Kronecker graph with 128M vertices and a degree of 16, then calls
breadth first search (BFS) for 64 randomly-chosen keys. 

The SPEC benchmarks \bwaves and \roms
are memory-intensive and were used by prior work~\cite{lee2023memtis,song2025hybridtier}.


%

\paragraph{Workloads}
One difference from previous evaluations is that we ran the applications on
larger datasets.
We chose
workloads for our applications such that the working set size (WSS) is
approximately 4GB, to
reflect real-world setups (such as production cloud environments running
multiple tenants~\cite{zhong2024managing}).
We define the WSS as the
average number of pages accessed in each 50ms period, times the page size.
We used a 50ms period because most previous tiered memory systems perform
migrations in the background every 50ms, \emph{independent} of cooling and
threshold adaptation.

We ran each of our experiments twice:  once provisioning the fast tier
that approximately matches the WSS, and another time providing 16GB fast tier size, about
4\myx of the WSS.


\subsection{Categorizing Memory Behaviors}
\label{sec:application-memory-behavior}
We categorize the
applications along two dimensions: how they allocate memory and how they
access memory over time. \Cref{fig:subject-applications} shows these categories for our applications.

\paragraph{Size of allocations}
Some applications create and manage memory primarily in objects that are larger
than a page (4KB, 2MB or 1GB) (e.g., GAP~\cite{beamer2015gap},
FlexKVS~\cite{kaufmann2016high}). Other applications create millions of objects that are smaller than a page (e.g., Btree~\cite{achermann2020mitosis} and Silo~\cite{tu2013speedy}). Some application have a mixed distribution of objects (e.g., \roms). 

\paragraph{Changes in hot data}
The hot data may remain stable and static throughout the execution
(e.g., XSBench~\cite{tramm2014xsbench}), or the hot data may exhibit phase
changes over time (e.g., GAP~\cite{beamer2015gap},
FlexKVS~\cite{kaufmann2016high}).

\subsection{Existing systems perform poorly for applications with small allocations}
\label{sec:analysis:allocation}

\begin{figure}
  \centering
\includegraphics[width=0.47\textwidth]{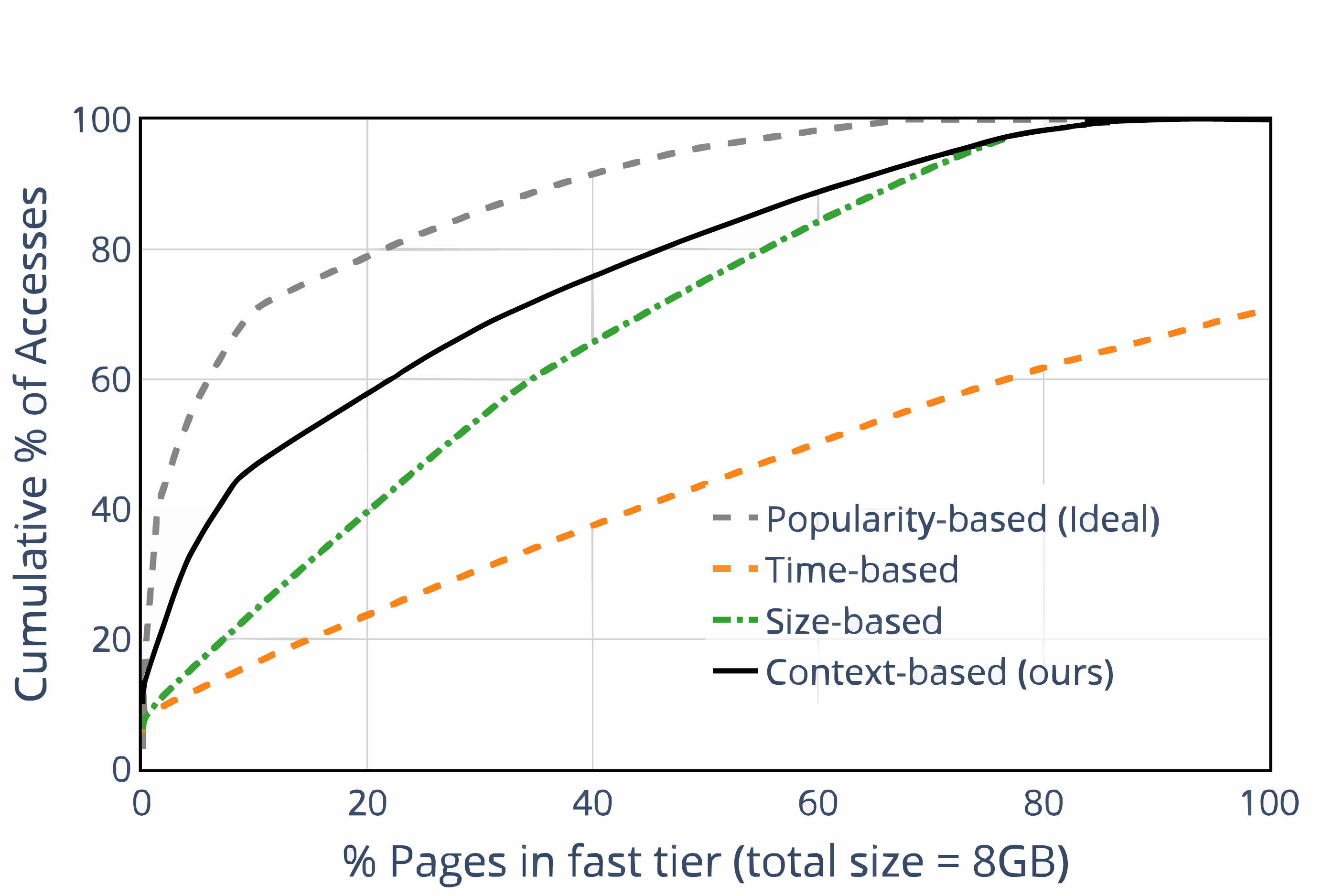}
\mycaption{Impact of object grouping}{This figure shows the CDF of memory
  accesses to pages for a read-only Btree workload, based on different object grouping
  strategies. A hypothetical popularity-based allocator packs all the
  popular objects together spatially, achieves a higher number of memory
  accesses in a fewer number of pages, followed by context-based grouping (\sref{sec:design:allocation}). 
}

\label{fig:mot-obj-grouping}
\end{figure}

Applications manage memory via heap-based objects that are
arbitrarily-sized, but tiered-memory systems track and migrate memory in pages that are 4KB, 2MB, or 1GB in size.

When allocating objects, memory allocators try to avoid fragmentation by
placing objects contiguously based on time of allocation
(e.g., glibc-malloc~\cite{gnumalloc}) or based on their size
(e.g., JeMalloc~\cite{argyroudis2012exploiting},
TCMalloc~\cite{zhou2024characterizing}),
and try to reuse space to avoid holes in the address space. In the presence
of skewed workloads and small objects, both of which are common to
datacenter applications~\cite{memcached2003,corbett2012spanner,atikoglu2012workload,yang2021large,pawar2025objectier}, this leads to popular objects being
scattered across a large set of pages.

In order to quantify the impact of allocators on page hotness,
\Cref{fig:mot-obj-grouping} shows the CDF curve of memory accesses across
pages for a
read-only \btree workload, for three allocation strategies:  using glibc-malloc~\cite{gnumalloc} (time-based grouping),
JeMalloc~\cite{argyroudis2012exploiting} (size-based grouping), and a
hypothetical popularity-based allocator.
The popularity-based allocator sorts objects based on their popularity,
then places them linearly in the virtual address space of the application.
(We computed
the popularity of objects
by intercepting all \sysmalloc and \sysfree calls  and 
by sampling memory accesses using PEBS, and mapping these to objects based on their
virtual address.)
\Cref{fig:mot-obj-grouping} shows that the allocators show significantly
different page hotness, and glibc-malloc and JeMalloc require $\sim$60\% and $\sim$30\% of the pages repectively in the fast tier for capturing 50\% of total
memory accesses, compared to the popularity-based allocator which requires $\sim$5\% of the pages. 
\Jenga's allocator uses a new grouping
strategy called \textit{context-based grouping}
(\sref{sec:design:allocation}) to group similar objects together for each
application.  It aims to approximate the performance of the hypothetical
popularity-based allocator.
\subsection{Existing systems perform poorly for applications with changes in hot data}

\label{sec:analysis:migration}
A high-performance tiered-memory system must promptly react to changes in
the hot region during application execution, to ensure that pages that
are no longer hot do not consume precious fast-tier memory forcing the newly
hot pages to remain in the capacity tier. We now analyze how previous tiered-memory systems adapt to changes in the hot data. 

\begin{figure}
  \centering
\includegraphics[width=0.48\textwidth]{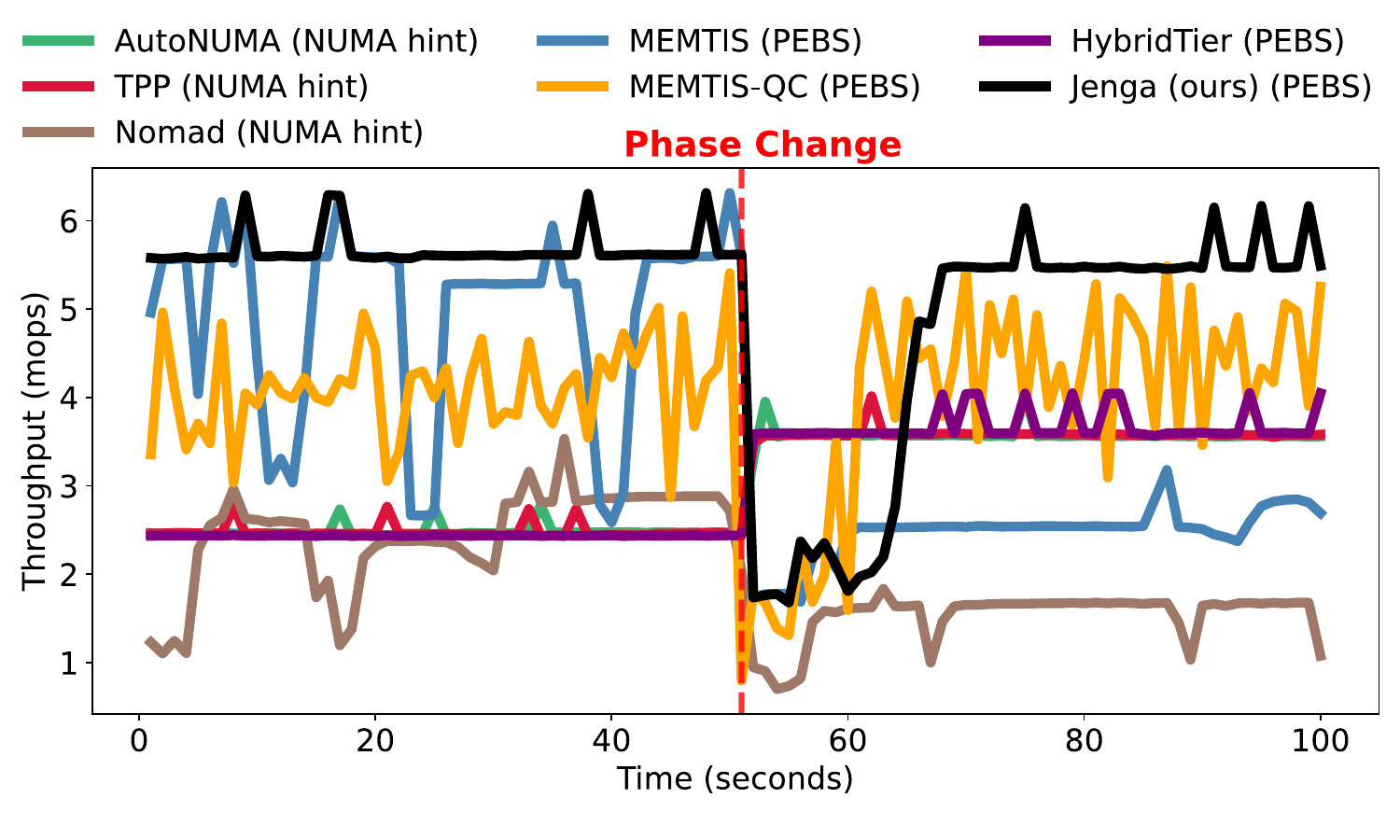}
\mycaption{FlexKVS instantaneous throughput}{Instantaneous performance for
  FlexKVS across tiered memory systems. There is a
  phase change at 50 seconds. No previous system achieves both high
  performance in a stable phase and timely adaptation to phase changes.
}
\label{fig:eval-flexkvs}
\label{fig:mot-flexkvs}
\end{figure}


\textbf{PEBS-based systems.}
To capture hot data changes, systems such as \memtis~\cite{lee2023memtis}
use periodic, infrequent cooling that captures the exponential moving
average (EMA) with a decay factor of $\frac{1}{2}$ for the page access counters
(\sref{sec:background:hotness:pebs}). The infrequent nature of cooling causes \memtis to
react slowly to changes in the hot data.
Our workload for the FlexKVS application exhibits a phase change in the
middle of the execution (\cref{fig:mot-flexkvs}).  \memtis never adapts to
the change in hot data.

Reducing the \coolinginterval (shown in \cref{fig:mot-flexkvs} as \memtisqc
for ``quick cooling'') allows quick adaptation to
working set changes, but it also increases instability within \memtis,
causing it to perform poorly and incur many migrations even in
a stable phase. The small \coolinginterval combined with exponential decay in each cooling event causes the pages to move dramatically, leading to thrashing. 



\hybridtier~\cite{song2025hybridtier} tries to overcome this limitation of \memtis by using
two \coolinginterval{s}: a large frequency interval of 80M samples to capture
long-term trends and a short momentum threshold of 500K samples to promptly
capture hot data changes. However, the use of rigid thresholds for determining page hotness for the momentum threshold (3 accesses to a page), along with a low PEBS sampling frequency for its low-resolution access counters causes \hybridtier to perform poorly in both the phases, as shown in \cref{fig:mot-flexkvs}.


\textbf{NUMA-hinting systems.}
NUMA-hinting systems do not capture long-term trends, but rather only
capture recent accesses. Such systems mis-identify cold pages
with one (or two) recent accesses as hot. This behavior is visible in
\cref{fig:mot-flexkvs} where Nomad and TPP, the state-of-the-art
NUMA-hinting systems, suffer from high latency even in a stable phase.

\reviewercomment{I understand your description in Section 3.4, identifying the downside \Memtis default parameters and your choice of parameters for \Memtis-QC, but have to admit did not entirely understand why it is not possible to align the intervals entirely and to make migration decisions at interval boundaries. Perhaps that can be better explained.}

\reviewercomment{I wasn't sure why the problem occurs because cooling happens for all the pages so the hotness, which is a relative measure, should still be retained. There was not a clear discussion on why hotness suddenly disappears with the cooling phase.}

\reviewercomment{
Figures 6 and 7 contrast Jenga with \Memtis. You seem to generalize \Memtis's behavior, but there is no convincing discussion that \Memtis's behavior (e.g., sawtooth hotness tracking) is general to all PEBS based approaches.
}

\subsection{Building a tiered-memory system}
\begin{figure}
  \centering
  \includegraphics[width=0.43\textwidth]{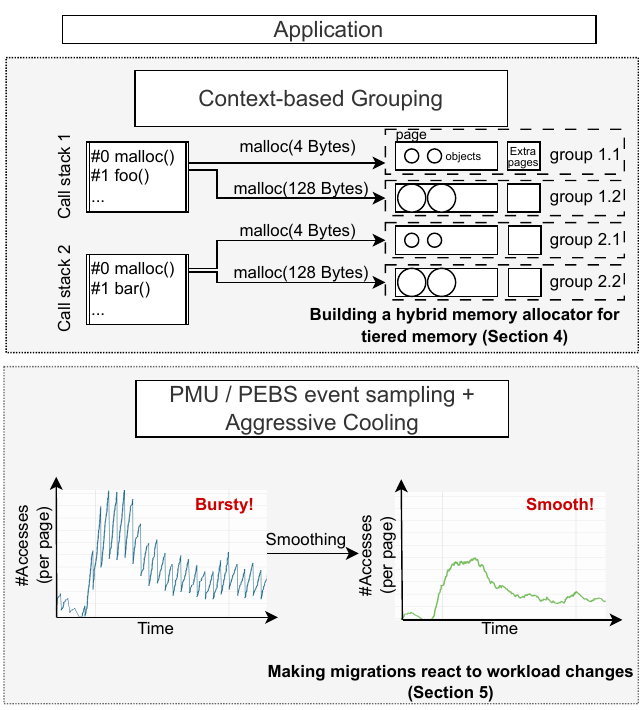}
\mycaption{Overview}{\Jenga uses a hybrid allocator for grouping objects (\cref{sec:design:allocation}) and smooth hotness tracking for adapting to workload changes while controlling migrations (\cref{sec:design:migration}).
} 
\label{fig:design-overview}
\end{figure}

This paper presents \Jenga, a tiered memory system that achieves high
performance for applications regardless of memory access behavior. \Jenga
differs from previous tiered memory systems in two ways.
It allocates objects so that co-hot objects are likely to be in the same
pages (\sref{sec:design:allocation}).
It tracks the hotness of each page in a smooth, rather than sawtooth,
manner, which enables timely migrations without thrashing (\sref{sec:design:migration}). \Cref{fig:design-overview} overviews the components of \Jenga.

\section{Hybrid Heap Memory Allocation}
\label{sec:design:allocation}

The allocation strategy is especially important when applications create
numerous small objects, as is common in datacenter applications. For
example, in Google applications, small objects ($<$4KB in size) account for
more than 90\% of all allocations, and objects smaller than 2MB in size
occupy $>$75\% of total allocated
memory~\cite{zhou2024characterizing}. Existing tiered memory systems rely
on off-the-shelf heap memory allocators that are designed for homogeneous
memory systems, which inadvertently scatter popular objects across numerous
pages, resulting in a poor hit rate, as discussed in
\sref{sec:analysis:allocation}.

\reviewercomment{You say creation of numerous small objects is "common" but give only one citation, which is a bit weak to be considered common. The small object size of <2MB seems a bit large. If you are going to consolidate small objects into pages, they would need to be much smaller. (Are you considering only huge pages?)}


\subsection{Goals for an allocator for tiered memory}

An allocator for tiered memory should enable the migration system to
perform effective page migrations.  First, the allocator should achieve \emph{object affinity}, \ie it should pack
objects of similar hotness into the same page. Second, the
allocator should reduce internal fragmentation, so that all the objects can
be placed in a minimal set of pages.


\subsubsection{Existing memory allocators are insufficient}
\label{sec:design:allocation:existing}
Existing memory allocators for homogeneous memory reduce memory fragmentation via size-based grouping of objects, but do not take into account the popularity of objects within a page, and thus lead to scattering of hot objects across pages.

Recent works such as MaPHeA~\cite{oh2021maphea} and
SOAR~\cite{liu2025tiered} build memory allocators for tiered
memory. However, these allocators do not work with a migration system, and
do not support running new applications and workloads without profiling
beforehand. These allocators rank objects based on their popularity using
profiling of workloads, and then place objects in the appropriate memory
tiers based on their popularity. They do not specialize on grouping of objects
within pages.  Moreover, they require a profiling run.


\begin{figure}
  \centering
\includegraphics[width=0.45\textwidth]{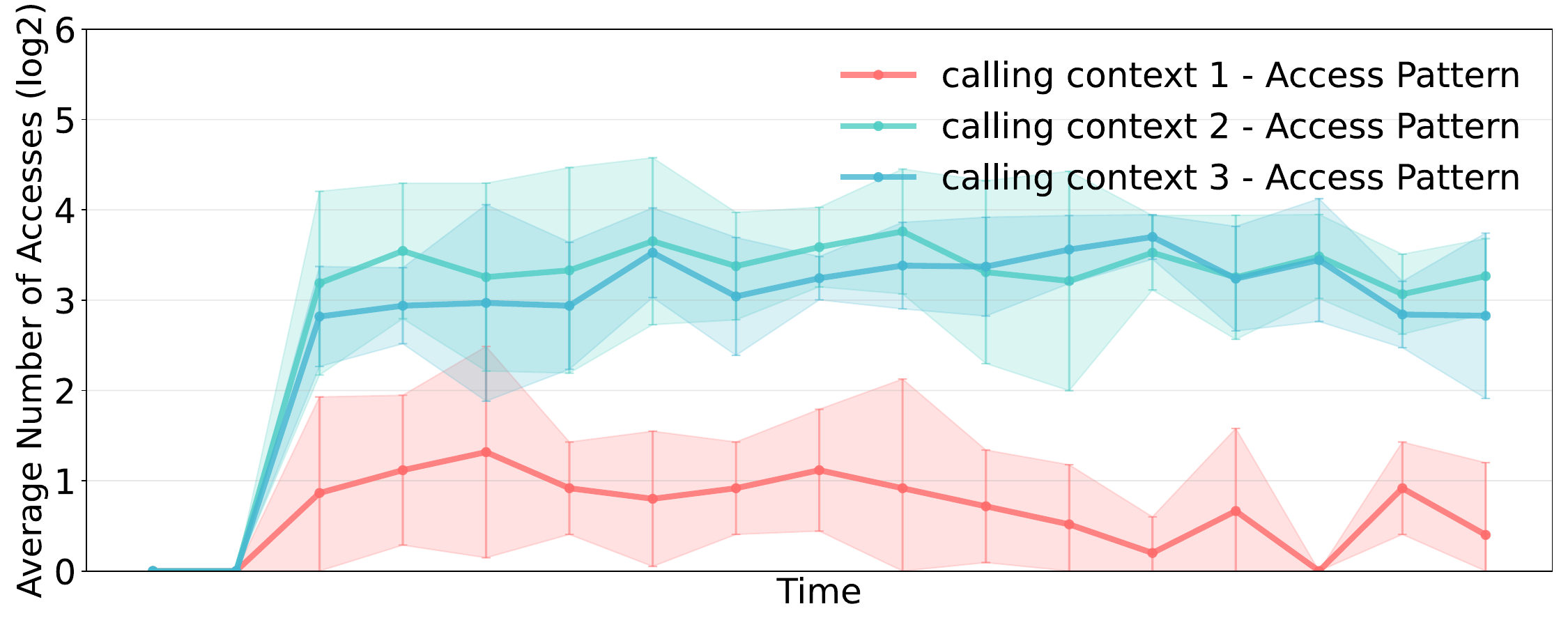}
\mycaption{Calling context affinity}{This figure shows the access pattern
  of different objects of the same size, but allocated from different
  calling contexts for the \btree application. Objects belonging to calling
  context 1 have a different popularity compared to objects belonging to
  calling contexts 2 \& 3.
}


\label{fig:design-call-chain-affinity}
\end{figure}


\subsection{Achieving better grouping}
We run the \btree application, and study the access patterns of different randomly-selected objects that share a calling context and allocation size, and compare it with objects that are of the same size, but belong to different calling contexts. \Cref{fig:design-call-chain-affinity} shows that objects with the same calling context and size exhibit similar access patterns compared to objects that do not share a calling context. Based on this observation, we introduce a new form of grouping objects called \emph{context-based} grouping that achieves object affinity by grouping objects into pages based on their calling context during allocation, as well as their size. Context-based grouping aims to efficiently pack objects within pages, instead of focusing on the placement of objects in a particular memory tier.

\subsection{Building a hybrid memory allocator}

In order to achieve context-based object grouping, the memory allocator
must support grouping of objects based on \emph{both} calling context and
size. Grouping by calling context colocates objects with similar access
patterns, while grouping by size helps to reduce memory
fragmentation. \Jenga's hybrid memory allocator accounts for both. \Jenga's
allocator is implemented as a library that links to the application,
intercepts all the memory allocation and free calls, and applies
context-based grouping for the objects at allocation
time. 

\Jenga's allocator first partitions the entire virtual address space of the
heap for an application into allocation regions. Each allocation region is
used to allocate objects that share a particular call stack. At the time of
allocation, \Jenga chooses an allocation region by hashing the return
addresses of all the functions in the backtrace, up to some depth.

Within each allocation region, \Jenga uses the JeMalloc allocator for
grouping objects based on size. JeMalloc implements hundreds of size-classes to avoid fragmentation; every object is assigned to its nearest size class, and is collocated with
other objects belonging to that size class.

\Jenga's allocator uses two hardcoded parameters.
(1) It uses 32 allocation regions.  This does not increase fragmentation
since the virtual address space is very large, but empirically it is large
enough to prevent collisions between dissimilar contexts.
(2) It observes the backtrace to a depth of 10.  Empirically, this is an
acceptable tradeoff between the overhead of traversing stack frames and the
ability to differentiate objects. \Cref{sec:eval:sensitivity:stack} studies
the sensitivity of this parameter.

\section{Adaptive Hotness Tracking}
\label{sec:design:migration}


\begin{figure}
\includegraphics[width=0.45\textwidth]{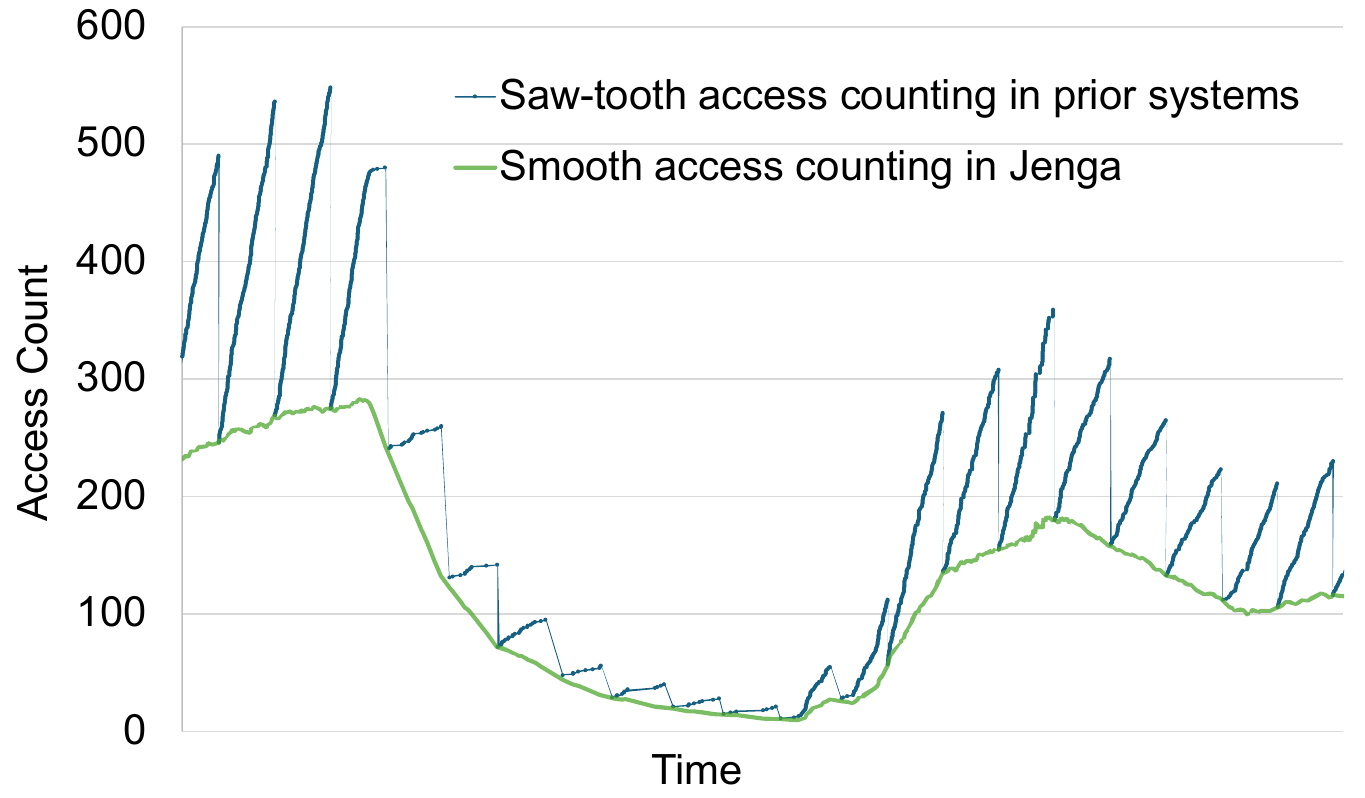}
\mycaption{Access counters of a page in \GAPBSBC}
{Access counters in prior systems (blue) monotonically increase until they are halved
  when the cooling period ends. \Jenga's
  smooth access counting (green) is more stable while still
  capturing changes within and between cooling periods.
}

\label{fig:est-plot}
\end{figure}


Previous PEBS-based tiered-memory systems~\cite{raybuck2021hemem,lee2023memtis,song2025hybridtier,ren2024mtm} adapt to changing hot data via cooling that periodically
halves the access counter of each page.  Tweaking the hard-coded parameters of
this cooling strategy does not perform well across diverse applications
(\sref{sec:analysis:migration}). We now formally describe how previous
systems capture memory access trends for a page, followed by \Jenga's
method of capturing memory trends.

\subsection{Capturing memory access trends of a page in previous systems}
\label{sec:design:existing_cooling}
Let $A(p, t)$  be the total number of accesses to a page $p$ until time $t$.
This number monotonically increases throughout the lifetime of the page.




\paragraph{Page access counters decay in large steps, only during cooling}
Previous systems design cooling as a periodic process, in which the access
counters of a page asynchronously decay by a factor of $\frac{1}{2}$ at the
end of a \coolinginterval. This causes sudden drops in the hotness of a
page.

Let \|cp| be the length of each \coolinginterval, and $\|cpstart|(t) =
\lfloor{t / \|cp|}\rfloor \cdot \|cp|$ be the start of the cooling period
that contains $t$. Thus, $A(p, cpstart(t))$ is the number of accesses for
$p$ at the start of a cooling period that contains $t$. Let
$\|cpaccesses|(p, t) = A(p,t) - A(p, \|cpstart|(t))$ be the number of
accesses received by page $p$ so far during the current \coolinginterval.
Let $A_C(p, t)$ be the access counter of page $p$ at time $t$.
\begin{equation*}
A_C(p, t) =
\small
\begin{cases}
A_C(p, cpstart(t)) + cpaccesses(p, t) & \text{if } t \not\equiv_{\|cp|} 0 \\
\frac{1}{2} \times A_C(p, t-1) & \text{if } t \equiv_{\|cp|} 0
\end{cases}
\end{equation*}




The access counter of a page monotonically increases when $t
\not\equiv_{\|cp|} 0$ and decays significantly only when $t \equiv_{\|cp|}
0$. \Cref{fig:est-plot} shows the saw-tooth pattern in the value of $A_C(p, t)$.
The sawtooth pattern causes two problems.


\paragraph{Sensitivity to \coolinginterval}
Capturing page access trends in this way makes previous systems overly
sensitive to $\|cp|$, which is a configuration parameter that cannot be
dynamically changed during execution. Moreover, any given \|cp| is a poor
fit for some applications.
A small \|cp| responds to working set changes faster but is more vulnerable to thrashing, while a large \|cp|
leads to slow responsiveness to working set changes.
Tiered memory systems conservatively err on the side of using a relatively large $\|cp|$.

Using two \coolinginterval{s} instead of one (as done in \hybridtier~\cite{song2025hybridtier}) to capture long-term trends and short-term bursts does not solve this problem. This is because, the system still remains sensitive to the two \coolinginterval{s}. Moreover, the smaller \coolinginterval remains vulnerable to thrashing, and hence still needs to be large enough to prevent it. For example, \hybridtier keeps the smaller \coolinginterval (called momentum interval) to 500K samples, which is still not enough to capture bursts in some cases, as shown in \cref{fig:mot-flexkvs}.


\paragraph{Volatility in access counter}
The sudden halving of $A_C(p, t)$ for a frequently accessed page causes its
hotness to change frequently, leading to unnecessary migrations.

\Memtis tries to mitigate this problem by classifying one bin as \emph{warm} between the hot and cold bins, but we will show why it still suffers from thrashing. 
In their intended design, given that bin sizes are factors of 2 and
cooling uses a decay of $\frac{1}{2}$, a single cooling could downgrade a hot page at most into the warm bin, but never directly into the cold bin, which should have prevented saw-tooth-related thrashings. 
However, two pitfalls weaken the intended mitigation. 

First, as noted in \sref{sec:memtis-no-warm-bin}, \Memtis sometimes disables the
warm bin, classifying every page as either cold or hot.  Consequently, the saw-tooth of the access counter would frequently go across the threshold, leading to undesirable
migrations.

Second, the saw-tooth approach is overly sensitive to the accuracy of PEBS sampling. PEBS exhibits inherent sampling variability~\cite{yi2020precision} and its sampling frequency changes dynamically during execution based on CPU utilization~\cite{lee2023memtis,song2025hybridtier}.
In the saw-tooth approach, pages in a stable hot phase must accumulate sufficient accesses within each cooling period to advance to a higher bin before exponential decay returns them to their previous bin. Concurrently, periodic threshold adaptation periodically adjusts the hot and warm thresholds during each cooling period based on the distribution of pages across bins (\sref{sec:memtis-no-warm-bin}).
This design creates instability when pages with similar access patterns receive slightly different PEBS sample counts due to natural sampling variation. These small measurement differences cause similarly accessed pages to accumulate counts at different rates, temporarily placing them in different bins. When threshold adaptation occurs within the same cooling period, these pages—which should be classified identically—migrate between different memory tiers. This migration pattern creates thrashing as the system responds to threshold changes.
Consequently, the saw-tooth approach amplifies measurement noise into classification instability, causing the system to react to sampling artifacts rather than genuine changes in access patterns. 


\paragraph{Use of alternative access tracking strategies is expensive} Alternative strategies for tracking accesses such as maintaining sliding window averages or using continuous exponential decay of counters (common solutions in the context of TCP congestion control~\cite{ha2008cubic}) can avoid the saw-tooth access pattern. However, these methods require maintaining an array of data proportional in size to the number of pages or the length of the time window, which introduce prohibitive memory overhead, rendering them infeasible.

\subsection{Smooth cooling: preventing the sawtooth}

\paragraph{Conceptually approximating continuous exponential decay by reducing \coolinginterval \emph{and} decay factor}


To flatten the spikes, one solution would be to distribute the exponential decay onto smaller time intervals. This can be done by reducing the decay factor and the \coolinginterval at the same time.
In the limit, the \coolinginterval
could be one access and the decay factor would be close to 0, approaching a continuous exponential decay.

However, implementing this simply by tuning parameters is not feasible, because cooling also involves heavy-weight and global
bookkeeping tasks such as traversing the migration queues and updating the
page-classification histogram (as discussed in \sref{sec:Background}).
Those tasks are designed with the assumption of a large
$\|cp|$~\cite{lee2023memtis}.

\paragraph{Decoupled smooth cooling}
\Jenga solves the problem of saw-tooth access counts by decoupling access
counter decay from the \coolinginterval and its heavy-weight bookkeeping. \Jenga slightly decays page
counters on \emph{every} page access, not just when $t \equiv_{\|cp|} 0$.

This decoupling of the page decay from other heavy-weight bookkeeping
enables \Jenga to apply a regular and an arbitrarily small decay to the
access counter of a page.  That small decay can be described as a function of the
\coolinginterval, designed intentionally in a way that at the end of each cooling period \Jenga's page
counters are the same as $A_C$.
The access counter of a page in \Jenga is:
\begin{equation*}
\resizebox{\hsize}{!}{$
A_S(p, t) = (\frac{1}{2} \cdot \frac{t - \|cp|start|(t)}{\|cp|}) \times A_C(p, cpstart(t)) + \frac{1}{2}\cdot cpaccesses(p, t)
$}
\end{equation*}

This allows \Jenga's page access counts to be smooth instead of saw-tooth (see \cref{fig:est-plot}) that does not result in sudden page hotness changes for frequently accessed pages in a stable working set. Furthermore, tracking accesses in this manner also reduces the sensitivity towards the inaccuracies and variations in the PEBS sampling, as pages in a stable working set do not frequently move across bins and hence do not require the hot bin threshold to change. Thus, by decoupling page decay with $\|cp|$, using a small $\|cp|$ to capture working set changes, and using a small decay factor at every time step to avoid unnecessary migrations, \Jenga is able to effectively adapt to an application's memory access behavior. 

The smooth cooling algorithm in \Jenga leverages only two counters (instead of an array of counters
required for alternative approaches), namely $A_C(p, cpstart(t))$ and $cpaccesses(p, t)$, and computes $A_S$ on every memory access, thus incurring a minimal memory overhead compared to the existing saw-tooth approach of tracking accesses.

\section{Implementation}
\Jenga's implementation has two main components: the context-based allocator and the migration component.

\paragraph{Context-based allocator}
The context-based allocator in \Jenga builds on top of the Memkind
library~\cite{cantalupo2015memkind}.  The library can be linked to the
application, or can be loaded at run time using LD\_PRELOAD. 

\paragraph{Migration component}
\Jenga's migration component is implemented by modifying Linux v5.15.19. It
supports hugepages as well as base pages.  It adds an access counter to the
\texttt{struct page}, which adds 12 bytes of metadata for every 4KB
pages to compute the smooth access distribution. \Jenga uses the spare
fields in Linux's \texttt{compound\_page} struct to support
hugepages without consuming any extra memory. Thus, \Jenga's memory overhead is bounded by 0.3\%.

\Jenga uses a k-thread for sampling memory access events using PEBS
at an adaptable sample frequency such that the CPU utilization is kept
below 3\% of a single core. The sampling thread implements the smooth
access distribution.
\Jenga uses one thread per memory region for migrations. The promotion / demotion threads wake up periodically to perform migrations by traversing the respective queues. 


\section{Evaluation}
\label{sec:evaluation}

We evaluate \Jenga by addressing the following questions:
\begin{itemize}[noitemsep,leftmargin=*,topsep=0pt]
    \item Does \Jenga outperform other tiered memory systems on diverse applications and fast tier capacities? (\sref{sec:eval:applications})
    \item Can \Jenga reduce datacenter memory costs? (\sref{sec:eval:memory})
    \item How do the components of \Jenga contribute to performance?
      (\sref{sec:eval:ablation})
    \item What is the expected performance of \Jenga on  CXL? (\sref{sec:eval:cxl})

\end{itemize}


\paragraph{Hardware setup}
We evaluated \Jenga on a dual-socket server equipped with Intel Xeon Gold
6230 @2.1 GHz processors (40 cores). The capacity tier is Optane DCPMM, similar to prior works~\cite{lee2023memtis,raybuck2021hemem,duraisamy2023towards,ren2024mtm}. We use a single socket for our evaluations to avoid NUMA effects.
We provision our fast tier capacities to applications as a function of their working set size, as discussed in \sref{sec:subject-applications}.



\paragraph{Comparison tiered-memory systems}
We compared \Jenga to \numPreviousSystems state-of-the-art tiered-memory
systems: \Memtis~\cite{lee2023memtis}, TPP~\cite{maruf2023tpp},
AutoNUMA~\cite{autonuma}, Nomad~\cite{xiang2024nomad} and \hybridtier~\cite{song2025hybridtier}. \Memtis, \hybridtier and \Jenga use PEBS counters for access tracking. TPP and Nomad rely on Linux's
LRU policy for demoting cold pages in the fast tier and use NUMA hinting
faults combined with static thresholds to identify hot pages in the
capacity tier. AutoNUMA does not have a page demotion mechanism and uses
NUMA hinting faults to detect hot pages in the capacity tier. \Memtis and \Jenga dynamically adjust page hotness based on fast tier size. \hybridtier uses two \coolinginterval{s} to track historical memory accesses (called frequency threshold) and to capture bursts (called momentum threshold), and uses static threshold of 3 accesses to a page for the momentum threshold to determine page hotness.
To evaluate the impact of \Jenga's smooth hotness tracking
(\sref{sec:design:migration}), we compare \Jenga to \Memtis configured with
the same \coolinginterval (120K samples), referred to as \MemtisQC (for
``quick cooling''). In addition,
we disabled page splitting in both \Jenga and \Memtis due to observed
instability and high variance in \Memtis when this feature is enabled. All applications thus use page sizes determined by Linux's memory manager. According to our observations, the applications do not entirely use only base pages or hugepages, but rather contain a combination of base pages and hugepages.

\reviewercomment{I believe Colloid [SOSP 2024], which you do not mention, controls migration considering latency. Perhaps other approaches such as Colloid have already mitigated this problem. So it would have been more convincing to see a comparison with other schemes such as Colloid.}

We do not evaluate Colloid~\cite{vuppalapati2024tiered} and ALTO~\cite{liu2025tiered}, because they are orthogonal to \Jenga{}. They do not alter the access tracking or cooling mechanisms in tiered memory. Instead, Colloid focuses on placing pages in the appropriate memory tier based on memory contention at each memory tier. ALTO uses LLC stalls rather than cache misses for computing page hotness. Both of those can be combined with any tiered memory system, including \Jenga, to enhance performance.  

\subsection{Applications}
\label{sec:eval:applications}

We ran a total of 10 applications whose descriptions and workloads are explained in \sref{sec:subject-applications}. We evaluated each application with a fast tier capacity that matches its Working Set Size (8GB for \btree and Silo, 4GB for the rest). We also evaluated each application with a larger and more conservative provisioning of the fast tier to 16GB. We ran each application 3 times and report the median speed-up relative to \Jenga.

\subsubsection{Btree: key--value store.}
\Cref{fig:eval-heatmap-apps} shows the access heatmap of \btree using
\Jenga's context-based allocation, which shows clearly distinguishable hot
and cold memory regions (dark rectangles). 

\Cref{fig:eval-combined-perf} (a) shows the performance when the fast
tier capacity is 8GB. \Jenga achieves the best performance, outperforming other systems by $>$\jengaVsMemtisEightGB\%{} due to context-based grouping in
\Jenga. All other systems use the default \glibc allocator and are unable
to consolidate hot objects into pages. \Jenga's hit rate is 83\% while
\Memtis's (the next best system) is 65\%. \Cref{fig:eval-combined-perf} (a) also shows the performance at a fast tier capacity of 16GB. 
Even at 16GB, the other systems are not able to fit hot data in the fast tier.

\subsubsection{Silo: In-memory database.}

\Cref{fig:eval-combined-perf} (b) shows the performance at fast tier capacities of 8GB and 16GB. \Jenga achieves the best performance at both fast tier capacities due to smart object grouping and adaptive hotness tracking. The gains are better for smaller fast tier due to a higher difference in hit rate compared to other systems.

\subsubsection{Graph Processing Benchmarks.}

\Cref{fig:eval-combined-perf} (c, d, j) show the performance for \GAPBSPR, \gapbsbc and Graph500 performing breadth-first-search on all tiered memory systems. Overall, \Jenga performs the best, equaling or surpassing \MemtisQC as the next best system for a fast tier capacity of 4GB, and \Memtis, when the fast tier capacity is 16GB. 

With a 4GB fast tier capacity, \Memtis and \hybridtier perform poorly compared to \Jenga due to a poor hit rate. On the other hand, with a 16GB fast tier capacity, \MemtisQC performs poorly as it suffers from a high number of migrations due to thrashing. 
\Jenga{'}s adaptive hotness tracking is able to perform timely migrations in reaction to phase changes, while avoiding thrashing, thus achieving the best performance.
Finally, the NUMA-hinting systems are not able to to identify the hot set accurately, and thus suffer from a poor hit rate, compared to the PEBS-based systems.

\subsubsection{XSBench: High Performance Computing.}
\begin{figure}
  \centering
\includegraphics[width=0.48\textwidth]{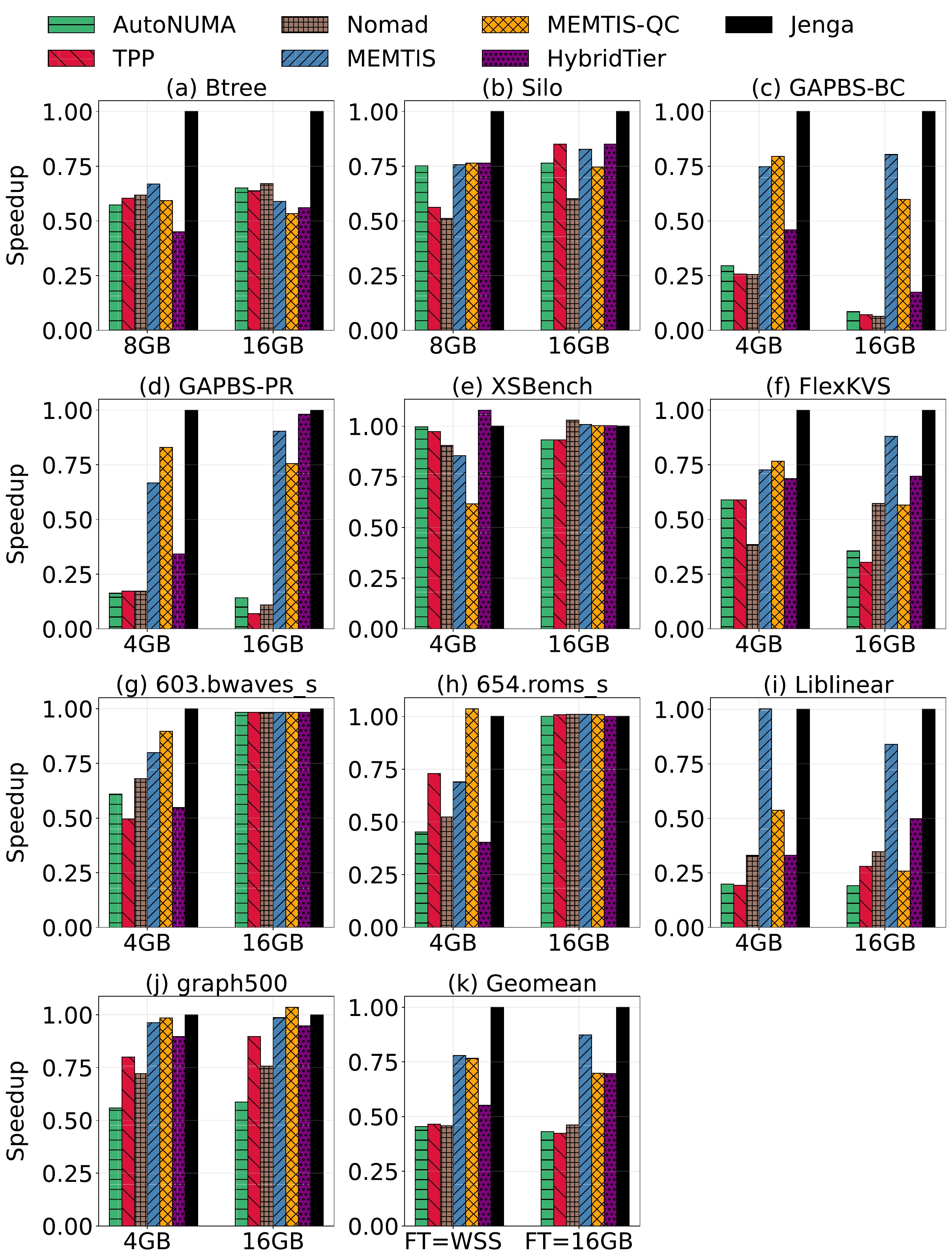}
\mycaption{Application Performance}{
  \Jenga outperforms
  the second best tiered memory system by \jengaVsSecondBestFastTierEqualsWSS\% when the fast tier
  equals WSS (4GB or 8GB) and by \jengaVsSecondBestFastTierLargerThanWSS\% when the fast tier is
  significantly larger than WSS (16GB). 
  }

\label{fig:eval-combined-perf}
\end{figure}

 \Cref{fig:eval-combined-perf} (e) shows the performance of tiered memory systems on XSBench. XSBench contains a highly skewed access pattern where the initial set of allocated pages remains hot for the duration of the workload. While all tiered memory systems achieve a high hit rate $>$80\%, the performance difference primarily comes from the initial placement of pages. The tiered memory systems that do not migrate out the initial pages achieve the best performance, as they do not require any migrations when the workload starts. As a result of this, \hybridtier achieves the best performance with no migrations. \Jenga performs 7\% worse compared to XSBench, because of having to migrate the hot data in the fast tier after the workload starts. \memtisqc suffers from instability due to its abrupt and frequent cooling, resulting in the worst performance.

 At 16GB, all tiered memory systems achieve a similar performance, as the total number of allocated pages in XSBench is smaller than the fast tier size, leading to a hit rate of 100\% in all the systems with no migrations.
 

\subsubsection{FlexKVS: in-memory caching.}
 \Jenga outperforms all other tiered-memory systems in FlexKVS by $>$30\% due to its adaptive hotness tracking. \sref{sec:eval:flexkvs} details the instantaneous performance of FlexKVS.

\begin{figure}
  \centering
\includegraphics[width=0.45\textwidth]{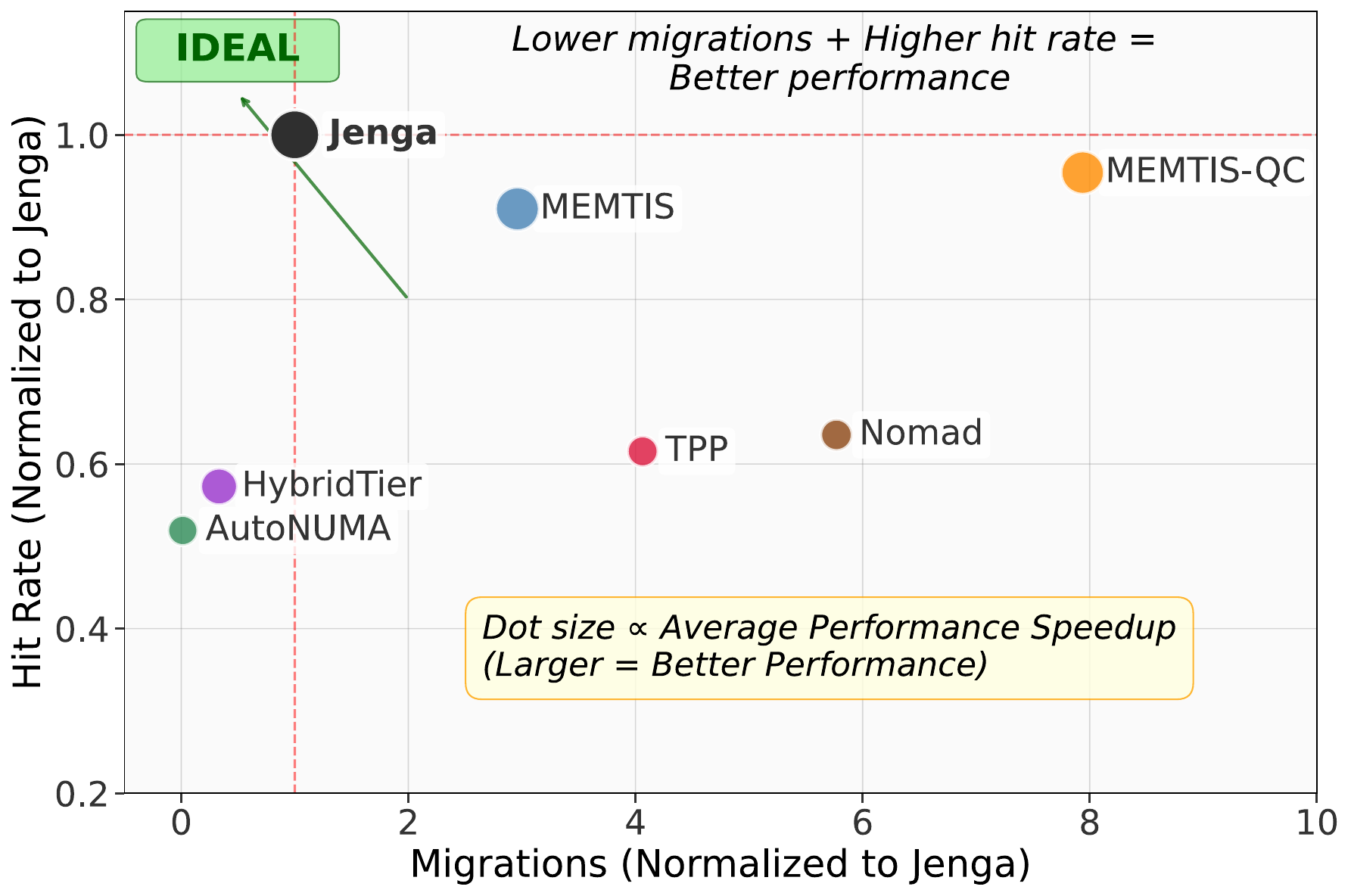}
\mycaption{Average Hit rate and migrations when fast tier size = WSS}{\Jenga gets the highest hit rate among all systems on average across applications, and incurs fewer migrations compared to systems that have a comparable hit rate. 
}

\label{fig:eval-hitrate-migs}
\end{figure}


\subsubsection{SPEC CPU 2017: \bwaves and \roms}

At a 4GB fast tier capacity, \Jenga outperforms all other systems on \bwaves, while performing second-best (5\% worse compared to \MemtisQC) in \roms, as shown in \Cref{fig:eval-combined-perf} (g, h). The main reason for the worse performance is that \roms exhibits sudden and dramatic changes in the hot data (visible via the checkered access pattern in \cref{fig:eval-heatmap-apps}), which aligns well with the cooling interval in \MemtisQC. The smoothed decay in \Jenga is unable to perform abrupt cooling, and thus achieves a lower hit rate compared to \MemtisQC. 
At 16GB fast tier capacity, all the systems perform equally, with a 100\% hit rate and no migrations.

\subsubsection{Liblinear: Machine Learning}

\Cref{fig:eval-combined-perf} (i) shows the performance at a fast tier capacity of 4GB and 16GB. Liblinear exhibits hot data changes at the granularity of 10M accesses, which is large enough for \Memtis{'s} large cooling interval, which performs similar to \Jenga. The adaptive hotness tracking of \Jenga shines in this application leading to 2\myx higher performance due to fewer migrations, compared to \MemtisQC with the same cooling interval. 
At a fast tier capacity of 16GB, \Jenga outperforms \Memtis due to fewer migrations at the same hit rate.






\subsubsection{Summary}
\Jenga overall results in being the superior tiered-memory system across different fast tier capacities and memory access patterns. 
It achieves \jengaVsSecondBestFastTierEqualsWSS\% better performance on average compared to the next best system when the fast tier matches
the working set size (WSS), and \jengaVsSecondBestFastTierLargerThanWSS\% when the fast tier is up to 4\myx larger than the WSS. While \MemtisQC performs second best when fast tier = WSS, its performance drops with a larger fast tier, due to excessive migrations. The rigid and large \coolinginterval{s} in \hybridtier along with low-frequency PEBS sampling inhibits it from promptly adapting to the working set. NUMA-hinting systems (Nomad, TPP, AutoNUMA) generally
underperform compared to PEBS-tracking-based systems due to inaccurate
access tracking.

\subsection{Performance on write-heavy workloads}
\Jenga employs a new hybrid memory allocator that groups objects based on the calling context and size, to approximate popularity-based grouping. While this benefits read-heavy workloads due to better packing of objects within pages, \Jenga incurs a modest overhead for write-heavy workloads that involve allocations in the critical path. In the worst case, for a workload that purely contains writes involving millions of small object creations, which occurs during loading keys in the in-memory index for \btree and in-memory database for Silo, \sysname increases the index loading time from 56 seconds to 59 seconds for \btree, while maintaining the same loading time of 260 seconds for Silo, compared to the others.

\begin{figure}
  \centering
\includegraphics[width=0.46\textwidth]{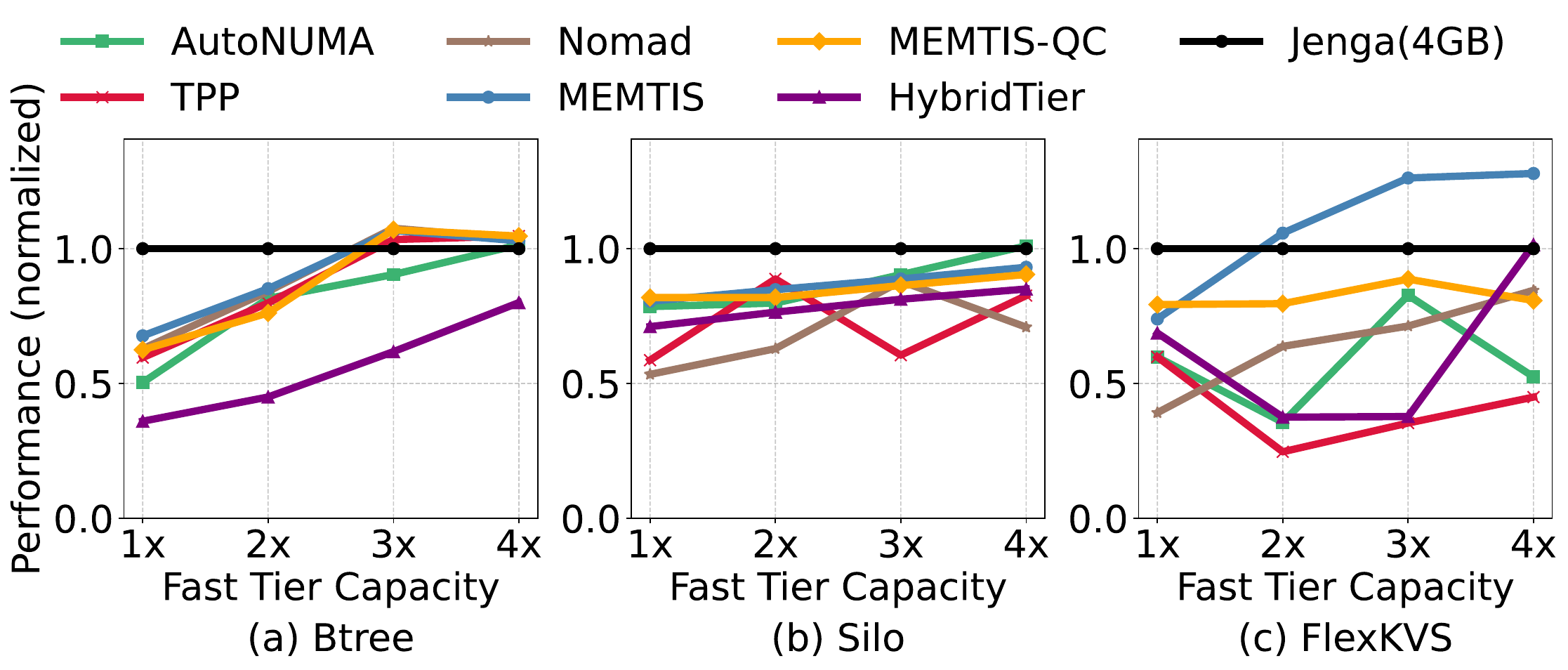}
\mycaption{Memory savings with \Jenga}{\Jenga can match or outperform other systems with much higher fast tier capacities, exhibiting significant memory savings.}
\label{fig:eval-memory-savings}
\end{figure}


\subsection{Memory savings with \Jenga}
\label{sec:eval:memory}
We evaluate the memory savings that are possible using \Jenga on three applications. \Cref{fig:eval-memory-savings} shows the performance of other tiered memory systems at different fast tier capacities, relative to \Jenga at fast tier = WSS. We show three applications (Btree, Silo and FlexKVS), because other systems with additional fast tier memory are able to achieve a better throughput than \Jenga configured with a 4GB fast tier capacity. \Jenga, thanks to its context-based allocations, is able to achieve memory savings of 2.5$\times$ in Btree and $4\times$ in Silo. In the case of FlexKVS with a changing hot data region, \Jenga{'s} adaptive hotness tracking allows it to achieve memory savings of almost 2$\times$ compared to the next best tiered memory system.

\subsection{Understanding the Performance of \Jenga}
\label{sec:eval:ablation}

\subsubsection{\Jenga achieves the best trade-off between hit-rate and migrations.}
\label{sec:eval:ablation:hitrate}
The performance of a tiered memory system depends on the percentage of memory accesses that it can serve from the fast tier (i.e. hit rate), and the number of migrations incurred to achieve a particular hit rate. \Cref{fig:eval-hitrate-migs} shows the hit rate against the number of migrations incurred by \Jenga relative to other systems on average across all our applications. \Jenga achieves the highest hit rate across all tiered-memory systems. While \MemtisQC achieves the second highest hit rate, it incurs $\sim$8\myx more migrations on average to achieve that hit rate.

\begin{figure}
  \centering
\includegraphics[width=0.48\textwidth]{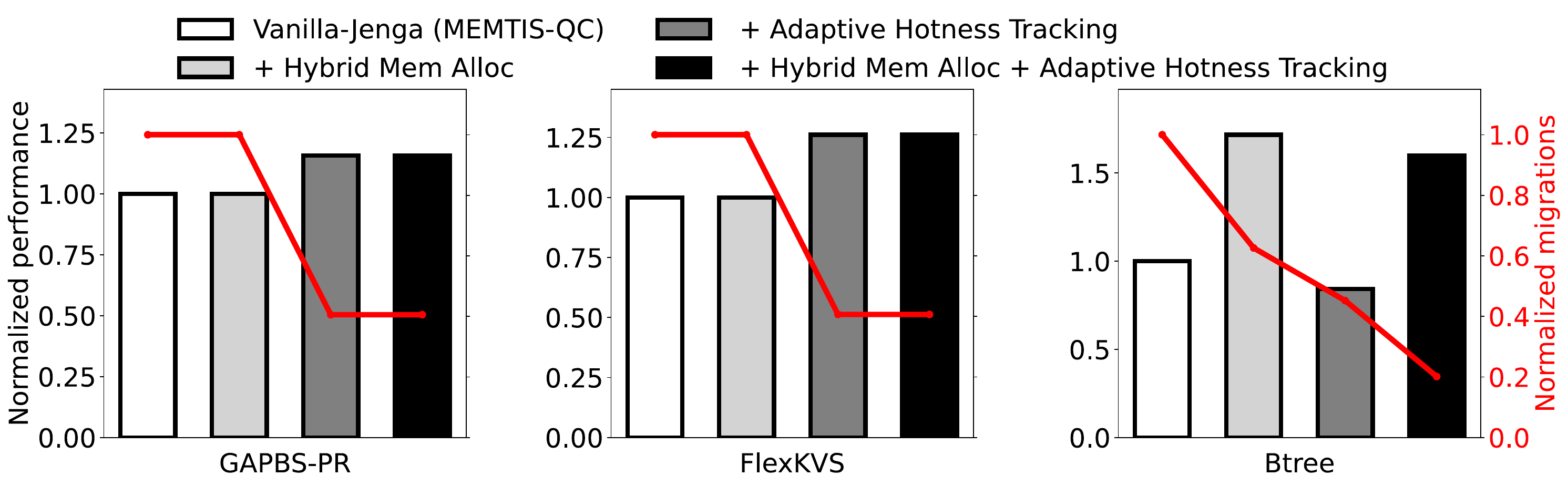}
\mycaption{Performance Ablation}{Adaptive hotness tracking and hybrid memory allocation in \Jenga reduces migrations and improves performance across diverse applications.}


\label{fig:eval-jenga-ablation}
\end{figure}


\subsubsection{Breaking down the performance of \Jenga}
\label{sec:eval:ablation:components}
We perform an ablation study to show the performance impact of each of our contributions. In each application, we show the end-to-end performance along migrations. We use our baseline as \MemtisQC (which we call Vanilla-\Jenga), as it has the same cooling interval as \Jenga. We then apply both our techniques individually and in combination to show their impact on migrations and performance in \cref{fig:eval-jenga-ablation}. \Jenga's smooth page hotness tracking significantly reduces the number of migrations for applications with large objects such as FlexKVS and \gapbspr. When we add context-based allocations on top of this, the performance of \btree improves significantly, while further reducing the number of migrations. 
Hence, each of our techniques helps a separate category of applications significantly, while adding up to a tiered-memory system that is able to perform well across a diverse set of applications.

\subsubsection{\Jenga is able to adapt to working set changes without incurring unnecessary migrations.}
\label{sec:eval:flexkvs}
\Cref{fig:eval-flexkvs} compares \Jenga's instantaneous performance to other tiered-memory systems on FlexKVS that exhibits a significant working set change at around 50 secs. \Jenga is able to achieve both stable performance as well as timely reaction to working set changes, compared to other tiered memory systems. 

\subsection{Expected performance using real CXL}
\label{sec:eval:cxl}
We were unable to run our benchmarks on real CXL (Intel Agilex 7~\cite{agilex}) since we did not have access to high-capacity CXL. Moreover, Intel does not yet have the capability to use PEBS on uncore events issued to the CXL memory controller which hampers the memory access tracking in PEBS-based tiered memory systems including \Jenga. We qualitatively discuss our performance on CXL compared to other systems.

First, the latency of CXL is higher than that of Optane~\cite{sun2023demystifying}. With a read access latency of around 400 ns (compared to 300 ns in Optane), the performance of a tiered memory system would be affected even more significantly by the hit rate of a tiered memory system. Furthermore, prior works have shown that a page migration in CXL can require approximately 50 microseconds~\cite{sun2025m5}. Thus, a tiered memory system that minimizes migrations will perform significantly better than one that incurs more migrations. \Jenga achieves the best ratio of hit rate to migrations (\sref{sec:eval:ablation:hitrate}), and is expected to outperform other systems even more significantly when run on CXL.

Note that we did not additionally evaluate on CXL emulation using DRAM on a remote NUMA node due to brevity and the significantly lower latency of NUMA memory (in the order of 150 ns after frequency throttling). While \Jenga would still outperform other systems in emulation, its main advantages are reduced accesses to the capacity tier, and reduced migrations, both of which are exacerbated with a higher latency capacity tier (such as PCIe-attached CXL memory).

\subsection{Parameter sensitivity and overheads.}

\subsubsection{Sensitivity of cooling interval}
\label{eval:adaptive-mig-effect}
\begin{figure}
  \centering
\includegraphics[width=0.45\textwidth]{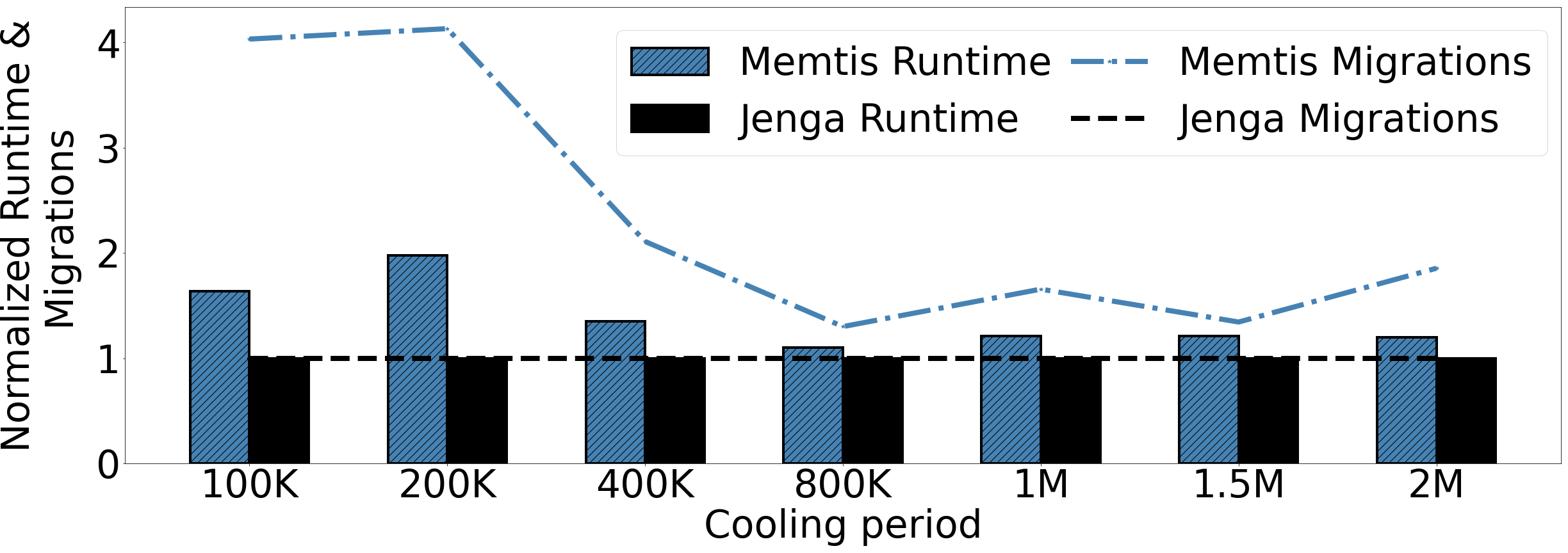}
\mycaption{Sensitivity to \coolinginterval}{\Jenga achieves robust performance and less sensitivity across \coolinginterval{s}. 
}
\label{fig:eval-cooling-impact}
\end{figure}

\Jenga reduces the overall sensitivity of the system towards the \coolinginterval with the help of smooth page hotness tracking (\sref{sec:design:migration}). We now evaluate \Jenga and \Memtis at various cooling intervals ranging from 100K samples to 2M samples. \cref{fig:eval-cooling-impact} shows the number of migrations and slowdown incurred by \Memtis relative to \Jenga at every \coolinginterval. \Memtis is significantly more sensitive towards the cooling interval. At small intervals, \Memtis incurs 4\myx more migrations than \Jenga. Moreover, even at \Memtis's default cooling interval of 2M samples, \Jenga outperforms \Memtis and incurs up to 2\myx lower number of migrations with a high hit rate.

\subsubsection{Sensitivity of context backtrace depth.}
\label{sec:eval:sensitivity:stack}
Context-based grouping requires finding the backtrace for allocations to
group objects based on the same call-stack. Choosing a depth for observing the call-stack at the time of allocation for grouping objects involves a trades off between fidelity and overhead. Observing allocations at a low call-stack depth could cause suboptimal groupings, since applications may use recursive calls or wrappers for their allocations. Conversely, traversing the call-stack at a high depth could make allocations slower and affect performance.

\begin{figure}
  \centering
\includegraphics[width=0.48\textwidth]{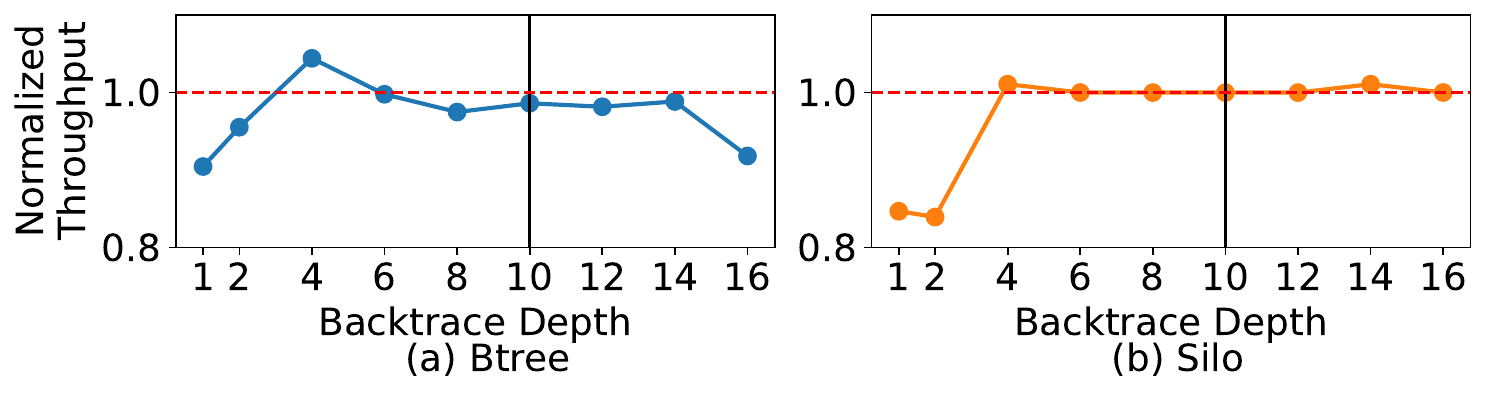}
\mycaption{Sensitivity to backtrace depth}{\Jenga is not sensitive to backtrace depths from 4--14. 
}

\label{fig:eval-stack-depth}
\end{figure}

We show the end-to-end performance of \Jenga using context-based grouping by observing the backtrace at different depths in \cref{fig:eval-stack-depth}. We observe that lower depths (1-3) are not able to group objects effectively, leading to poor performance. Higher depths, on the other hand, do not incur noticeable end-to-end costs in our applications. Since the end-to-end performance remains insensitive to the overheads of higher call stack depths (4--14), \Jenga uses a default call-stack depth of 10 across all applications. We find that this depth is large enough to group objects effectively. \Jenga's allocator requires 100 cycles on average to obtain the backtrace.

\section{Related Work}
\label{sec:Related-Work}

\noindent \textbf{Software-managed tiered memory systems.}
Prior software-managed tiered memory systems have explored various aspects of memory tiering that include access tracking mechanisms~\cite{raybuck2021hemem,lee2023memtis,maruf2023tpp,duraisamy2023towards,ren2024mtm} as well as page migrations~\cite{lee2023memtis,ren2024mtm,maruf2023tpp,duraisamy2023towards,yan2019nimble}. 
HeMem~\cite{raybuck2021hemem}, TPP~\cite{maruf2023tpp}, Nomad~\cite{xiang2024nomad}, TMTS~\cite{duraisamy2023towards}, Nimble~\cite{yan2019nimble},
AutoTiering~\cite{yang2017autotiering} and AutoNUMA~\cite{autonuma} use static thresholds to determine page hotness and cannot handle diverse applications.
Memtis~\cite{lee2023memtis} and MTM~\cite{ren2024mtm} are the closest systems to \Jenga that use histograms and fast tier capacities to guide page hotness detection and migrations. Recently proposed \hybridtier~\cite{song2025hybridtier} uses a combination of histogram and static thresholds for hotness detection.

None of the previous tiered memory systems evaluate the impact of allocations on tiered memory in combination with a migration system. 
Furthermore, they are all sensitive to static \coolinginterval{s}, and thus are only suitable for applications with static or gradually changing working sets.

\noindent \textbf{Hardware-managed memory tiering.}
Hardware solutions include Intel's memory mode for Optane and 2LM mode for CXL~,\cite{nvm} which use the fast tier as a direct-mapped cache. JohnnyCache~\cite{lepers2023johnny} uses OS page remapping to reduce cache collisions in 2LM. m5~\cite{sun2025m5} and NeoMem~\cite{zhou2024neomem} track capacity tier accesses via snooping mechanisms in the CXL FPGA.
However, hardware-managed solutions are limited to simple policies that don't generalize across diverse applications, and lack fine-grained control in multi-tenant datacenters. 

\noindent \textbf{Tiered Memory Systems in industry.}
 Google TMTS~\cite{duraisamy2023towards}, Microsoft Pond~\cite{li2023pond}, Facebook TPP~\cite{maruf2023tpp} offer industry's solution to memory tiering. Their primary aim is to reduce memory costs by partially replacing local DRAM with CXL/NVM, typically operating with >65\% fast tier size due to strict SLOs.

In contrast, \Jenga and similar systems (e.g., HeMem~\cite{raybuck2021hemem}, Memtis~\cite{lee2023memtis}, MTM~\cite{ren2024mtm}, Nomad~\cite{xiang2024nomad}, \hybridtier~\cite{song2025hybridtier}) focus on operating at significantly lower fast tier capacities, using sophisticated tracking and migration policies to maximize hit rates. These approaches are complementary, addressing different aspects of heterogeneous memory management. \Jenga specifically aims to maximize cost savings by minimizing expensive DRAM usage without compromising performance.

\noindent \textbf{Heap object management libraries.}
See \sref{sec:design:allocation:existing} for a discussion of existing homogeneous memory allocators and recently proposed allocators for tiered memory. None of the existing allocators for tiered memory~\cite{liu2025tiered,oh2021maphea} are designed to work with migrations, and they rely on profiling of applications for determining object popularity which is impractical.
The HML~\cite{miucin2018data} and Vam~\cite{feng2005locality} allocators attempt to achieve object affinity, similar to \Jenga's allocator. However, they either require changing applications~\cite{miucin2018data}, or use static policies that are not tailored to application memory behavior~\cite{feng2005locality}.




\section{Conclusion}
\label{sec:Conclusion}

\Jenga is a new tiered memory system that achieves
high performance for a variety of applications. \Jenga includes the allocation system as part of its design and performs adaptive hotness tracking that adapts to working set changes of an application without incurring unnecessary migrations. \sysname is able to outperform other tiered memory systems on a variety of applications with different allocation and memory-access patterns. We have made \sysname publicly
available at \url{https://github.com/efeslab/jenga}.


\bibliographystyle{abbrv}
\bibliography{main}


\end{document}